\documentclass{article}
\usepackage{amsmath}

\usepackage{PRIMEarxiv}

\usepackage[utf8]{inputenc} 
\usepackage[T1]{fontenc}    
\usepackage{hyperref}       
\usepackage{url}            
\usepackage{booktabs}       
\usepackage{amsfonts}       
\usepackage{nicefrac}       
\usepackage{microtype}      
\usepackage{lipsum}
\usepackage{fancyhdr}       
\usepackage{graphicx}       
\usepackage{siunitx}
\usepackage{placeins}
\usepackage{tikz}
\usetikzlibrary{positioning, shapes.geometric}
\usepackage{graphicx}
\usepackage{mwe}
\usepackage{caption}
\usepackage{subcaption}
\usepackage{float}
\usepackage{placeins}
\graphicspath{{media/}}     
\usepackage{makecell}
\usepackage{booktabs}
\usepackage{tabularx}
\usepackage{array}
\usepackage{multirow}
\usepackage{enumitem}

\title{A Time-Resolved Framework for Quantifying Neuronal Network State Transitions
\thanks{v1: This manuscript presents preliminary results and is a work in progress. The analysis and manuscript are subject to further revision before submission for peer review.} 

}

\author{
  Ilya Auslender\thanks{These authors contributed equally to this work.} \\
  Department of Physics, \\
  University of Trento, \\
  Italy \\
  \texttt{ilya.auslender@unitn.it} \\
   \And
  Yasaman Heydari\footnotemark[2] \\
    Department of Physics, \\
  University of Trento, \\
  Italy \\
  \texttt{seyedeh.heydari@unitn.it} \\
  \And
  Asiye Malko\c{c} \\
    Department of Cellular, Computational\\ and Integrative Biology, \\
  University of Trento, \\
  Italy \\
  \And
    Clara Zaccaria \\
    Department of Physics \\
  University of Trento, \\
  Italy \\
  \And
  Yuri Bozzi\\
  Center for Mind/ Brain Sciences,\\
  University of Trento,\\
  Italy\\
  \texttt{yuri.bozzi@unitn.it}
  \And
  Lorenzo Pavesi\\
  Department of Physics,\\
  University of Trento,\\
  Italy\\
   \texttt{lorenzo.pavesi@unitn.it}
}

\begin{document}
\maketitle

\begin{abstract}
Electrophysiological recordings provide powerful tools for characterizing the dynamics of neuronal networks. In vitro neuronal cultures offer a controlled setting for investigating network responses to diverse interventions, including pharmacological and optogenetic stimulation. However, conventional analyses often reduce electrophysiological activity to aggregate measures calculated over fixed temporal windows, providing only a static representation of the network state. Here, we introduce a comprehensive, multiparametric analytical framework designed to characterize both the temporal evolution of network activity and the transitional states induced by external interventions. The framework combines multiple network discriminators with a compositional analysis that resolves electrode-level responses into excited, inhibited, and unchanged states over successive temporal segments. We apply the framework to electrophysiological recordings obtained under pharmacological interventions with 4-aminopyridine (4-AP), optogenetic stimulation, and additional benchmark conditions. The results demonstrate that the proposed analysis can reveal subtle and transient responses that may be difficult to detect using conventional aggregate measures, including responses to low-dose pharmacological interventions. These findings highlight the value of temporally resolved and compositional analyses for characterizing heterogeneous neuronal network responses in vitro.
\end{abstract}


\section{Introduction}

\textit{In-vitro} platforms offer a highly controlled experimental environment that permits precise manipulation of biological variables, thereby providing distinct advantages over \textit{in-vivo} and clinical studies for isolating the mechanisms underlying neurological and neurodegenerative disorders \cite{osaki2018vitro}.
These systems offer a powerful framework for unraveling the complex cellular and molecular mechanisms that govern disease initiation and progression. \textit{In-vitro} models, in particular, enable precise interrogation of neuroinflammatory signaling networks that contribute to epileptogenesis and provide mechanistic insight into how tau pathology drives neuronal dysfunction and disease progression in Alzheimer’s disease \cite{pons2021dissecting,cerovic2023vitro,richard2021queuine}.

Furthermore, \textit{in-vitro} approaches enable real-time monitoring of how these pathogenic factors alter the functionality of the neuronal network \cite{jablonski2021experimental}. Techniques such as electrophysiological recordings and calcium imaging permit a quantitative assessment of neuronal activity, connectivity, and network propagation dynamics \cite{auslender2025decoding,zaccaria2026investigation}, which supports the identification of candidate targets for therapeutic intervention \cite{jablonski2021experimental,heydari2024decoding}.

These approaches are instrumental not only in the development and preclinical evaluation of targeted therapies, but also in assessing drug safety, toxicity, and dose–response relations prior to progression to \textit{in-vivo} or clinical studies \cite{heydari2024decoding}.

Electrophysiological approaches, such as microelectrode arrays (MEAs), provide a powerful means of interrogating the functional dynamics of neuronal cultures by capturing features such as firing rates, network synchrony, rhythmic activity, and excessive bursting, thereby enabling longitudinal assessment of neuronal network activity and organization \cite{pelkonen2021functional}. Despite these capabilities, conventional electrophysiological metrics such as mean firing rate (MFR) or burst indices provide only static snapshots of network activity, often obscuring the temporal dynamics of neuronal excitability. Firing rate, for instance, may indicate hyperexcitability but cannot reveal whether this state is transient, sustained, or oscillatory dynamics that are critical for understanding epilepsy models or synaptic plasticity \cite{zhu2025temporal}.
For instance, kainic acid, a widely used agent for inducing seizures in animal models, reliably produces well-characterized hyperexcitability \text{in-vivo} \cite{gisabella2012kainate}. Interestingly, \textit{in-vitro} studies of neuronal cultures have reported a reduction in firing rate following the same treatment, which could indicate the decreased neuronal excitability \cite{hyvarinen2019functional,konrad1999kainic}. This apparent discrepancy may arise from several factors. First, heterogeneity in the cellular composition of cultured populations may shape network-level responses and contribute to variability in the observed effects \cite{altschuler2010cellular}. Second, transient increases in neuronal activity immediately following drug application may be masked when analyses are based solely on average firing rates. In particular, an initial surge in activity occurring within the first few seconds may be diluted by subsequent reductions in neuronal firing, resulting in an overall mean that fails to capture these rapid and potentially biologically relevant dynamics.

In addition to the limitations of summary metrics such as MFR, which can obscure temporal changes in network behavior, another critical factor in \textit{in-vitro} studies is the cellular type composition of the culture \cite{pelkonen2021functional,crocco2025proper}. Evoked responses in neuronal culture networks are critically shaped by factors including neuronal subtype composition, astrocytic support, developmental stage, receptor distribution, and external stimuli \cite{crocco2025proper,ahtiainen2024astrocytes}. 
Existing approaches rarely integrate these temporal and compositional dimensions, constraining mechanistic interpretation and predictive power in preclinical studies. These limitations are particularly consequential when stimulus-evoked changes are modest, as subtle but biologically meaningful reorganization of activity patterns may occur without substantial changes in conventional aggregate measures. Resolving how network dynamics evolve in response to perturbation therefore requires analytical frameworks capable of capturing not only the magnitude, but also the temporal structure and patterning of the response.

To address these limitations, we introduce a comprehensive analytical framework that captures not only the magnitude of changes in network activity, but also their temporal evolution and persistence. We hypothesize that transient network dynamics are jointly shaped by the nature of the intervention and the cellular composition of neuronal populations—features that are poorly resolved by conventional aggregate measures. This limitation is particularly relevant for mild or subtle intervention-induced effects, for which changes in overall activity may be modest, while meaningful alterations in the temporal organization and persistence of network responses may nevertheless emerge.

Beyond the cumulative spiking characteristics of the cultures, our analysis examines the statistics of inter-spike intervals (ISIs) within distinct temporal windows. This approach provides additional information on changes in the temporal organization and dynamics of network activity, enabling the identification of intervention-induced effects that may not be apparent from aggregate measures alone. A central component of the framework is the characterization of the response composition of the culture, by quantifying the relative fractions of neuronal populations exhibiting increased activity, decreased activity, or no appreciable change in response to the intervention. This population-level characterization captures the heterogeneity of network responses and provides a means of resolving subtle effects that may be masked when activity is considered only at the level of the culture as a whole.

In this study, we applied the proposed framework to quantify the responses of neuronal cultures to diverse interventions and to characterize the subtle, transient dynamics of their activity over time. Our results demonstrate that these responses are influenced not only by the type of intervention, but also by the specific neuronal subtypes represented in the cultures. These findings highlight the importance of temporally resolved and cell-type-aware analyses for elucidating intervention-induced network dynamics and for improving the interpretation of \textit{in-vitro} models in mechanistic studies and preclinical evaluation.

\section{Method and Materials}
\label{sec:headings}

\subsection{Neuron Isolation: Techniques and Protocols for Neuronal Culture Preparation}
All procedures involving neuronal isolation were conducted in strict accordance with Italian Animal Welfare legislation (D.Lgs. 26/2014), which implements the Directive 2010/63/EU of the European Parliament and of the Council on the protection of animals used for scientific purposes. The experimental protocols were reviewed and approved by the competent local veterinary authorities and authorized by the Italian Ministry of Health.

\paragraph{Primary Neuronal Culture}
Pregnant mice at embryonic day 17–18 (E17–E18) were euthanized by CO$_2$ inhalation. Embryos were rapidly harvested, decapitated, and transferred to ice-cold dissection buffer composed of Hank’s Balanced Salt Solution (HBSS, 1×; Gibco) supplemented with D-(+)-glucose (1 mM; Sigma) and HEPES (1 M; Gibco). Following careful removal of the skin and skull, whole brains were promptly isolated.

The meninges were removed and the cerebral cortices dissected under clean conditions. Cortical tissue was enzymatically dissociated in 0.25\% Trypsin–EDTA (Gibco) for 20 min at 37~$^\circ$C. Digestion was quenched with high-glucose DMEM supplemented with 10\% fetal bovine serum (FBS) and penicillin–streptomycin (10,000 U/mL; all from Gibco). The suspension was gently triturated, filtered to remove debris, and cells were counted.

Cells were pelleted (1,900 rpm, 5 min), resuspended in Neurobasal medium supplemented with 10\% FBS and penicillin–streptomycin, and plated at $1.36 \times 10^5$ cells in 80 µL ($1.7 \times 10^3$ cells/µL) per poly-D-lysine/laminin-coated microelectrode array (MEA) chip. Cultures were allowed to adhere for 2 h under standard incubation conditions prior to further medium addition and subsequent experiments.

Following this incubation, the feeding medium was added, consisting of Neurobasal (Gibco), B27 Supplement (50$\times$, Gibco), GlutaMAX Supplement (100$\times$, Gibco), and Penicillin-Streptomycin. Half of the medium was replaced every 3--4 days.  

To ectopically express light-sensitive ion channels in neurons, cultures were transduced at 3 days in-vitro (DIV) using a viral vector encoding the candidate protein (pAAV-CaMKII-ArchT-GFP, Addgene \#PV2527). The viral titer was adjusted accordingly (Titer: $8.4 \times 10^{9}$ GC per chip). Cultures were incubated overnight, and GFP or YFP expression was evaluated approximately 7 days later to confirm successful expression of the optogenetic construct.

\paragraph{Mouse Embryonic Stem Cells}
We further applied our computational framework to additional datasets generated by an independent laboratory employing different recording systems. In these datasets, cells were prepared and characterized as previously described by Crocco et. al. \cite{crocco2025proper}. Neurons were derived from mouse embryonic stem cells (mESCs) and differentiated into dorsal telencephalic (DT) or ventral telencephalic (VT) lineages. The final experimental groups included pure DT, pure VT, and mixed cultures at DT:VT ratios of 80:20 and 50:50 \cite{crocco2025proper}.

\subsection{Electrophysiological Monitoring}
Electrical activity of neuronal cultures was recorded using the MEA2100-MiniSystem (Multi Channel Systems GmbH) at a sampling rate of 10–20 kHz. Neurons were cultured on titanium nitride (TiN) microelectrodes integrated on glass substrates with polyimide insulation and an internal reference electrode. The microelectrode array comprised an 8 × 8 configuration with 60 active electrodes (30 µm diameter, 200 µm interelectrode spacing).

Spontaneous network activity typically emerged between 9–11 days in-vitro (DIV), and recordings were performed between DIV 12 and 23, when stable and robust activity was observed \cite{chiappalone2006dissociated,potter2001new}. Prior to data acquisition, cultures were allowed to equilibrate for 30 min to ensure signal stability \cite{potter2001new}.

For each recording across the experimental groups, electrophysiological data were acquired at a sampling frequency of either 20 kHz or 10 kHz, independent of the recording platform employed. Action potentials (spikes) were identified using the Precise Timing–Spike Detection (PTSD) algorithm described by Maccione et. al. \cite{maccione2009novel}. This approach detects candidate events by identifying relative local extrema within a predefined peak-lifetime window and subsequently evaluating their peak-to-peak amplitude against an electrode-specific, noise-adaptive differential threshold. The detection threshold is computed individually for each electrode to account for variability in baseline noise levels. A refractory period constraint was imposed to prevent multiple detections of the same spike event, thereby preserving temporal precision and enhancing discrimination from background activity.

For all mESC-derived culture groups \cite{crocco2025proper}, electrophysiological activity was recorded using high-density CMOS-based 4096-channel microelectrode arrays (Accura, 3Brain). This platform enabled high-resolution characterization of network-wide spiking and bursting dynamics. Spike detection was carried out using the previously established PTSD protocol \cite{maccione2009novel}, ensuring analytical consistency across experimental conditions and datasets.

To avoid unreliable estimates arising from insufficient statistical sampling, electrodes exhibiting a firing rate below $0.1\,\mathrm{s}^{-1}$ were excluded from further analysis and classified as inactive. Entire recordings were excluded from the dataset if fewer than 10 electrodes met the activity criterion, ensuring sufficient sampling of network-level dynamics and reliable estimation of population statistics.

\subsection{Experimental Groups and Paradigms}

\paragraph{Primary cortical neuronal cultures} Primary cortical cultures were assigned to seven experimental conditions comprising pharmacological, optogenetic, and control groups.

\begin{enumerate}
    \item  \textbf{Pharmacological modulation:} Network excitability was manipulated using 4-aminopyridine (4-AP) at final concentrations of 10\,\textmu M and 100\,\textmu M \cite{pozzi2013rest,prestigio2021rest}. Vehicle controls received sterile distilled water, and medium controls received plain medium. All treatments were delivered in a volume of 1\,\textmu L added to 999\,\textmu L culture medium. Cultures were allowed to stabilize for 30\,min prior to recording. Baseline activity was recorded for 10\,min, followed by treatment application and an additional 30\,min recording period.
    
    \item \textbf{Optogenetic modulation:} For light-dependent perturbation, ArchT-expressing cultures were subjected to patterned green light stimulation (10\,mW\,mm$^{-2}$). Illumination was generated using an LED source, spatially patterned via a digital micromirror device (DLP), and delivered through a microscope optical path in wide-field shape \cite{zaccaria2026investigation,auslender2023integrated}. The stimulation protocol consisted of 15\,s light ON followed by 60\,s OFF cycles. Non-expressing cultures receiving identical illumination served as light controls \cite{han2011high}. 
    
    \item \textbf{Untreated controls:} Additional control groups included spontaneous cultures (no intervention).
\end{enumerate}

\paragraph{mESC-derived neuronal dataset (Benchmark)}

An independent dataset of mouse embryonic stem cell (mESC)-derived neuronal cultures was analyzed for methodological validation \cite{crocco2025proper}. Four gruops of culture compositions were included: pure DT, pure VT, DT:VT (80:20), and DT:VT (50:50).

Recordings consisted of a 5 minute baseline spontaneous activity period followed by bicuculline application and a 6\,min post-application recording interval. Bicuculline was used to transiently reduce inhibitory signaling as described in Ref. \cite{crocco2025proper}. 

See Table \ref{tab:experimental_paradigms} for details of the experimental groups and their usage in the experiments of this work.

\subsection{Analytical Approach}

The method presented in this work is based on quantifying changes in neuronal network activity by comparing a measured or derived observable (hereafter referred to as a discriminator) between two defined time windows. The framework requires the electrophysiological dataset to provide the following elements:
\begin{enumerate}
\item \textbf{Defined baseline window:} The baseline window corresponds to the period of recording preceding the application of the external intervention or stimulation. Changes in the activity of the culture are quantified relative to the baseline, providing a reference for assessing intervention-induced effects.

\item \textbf{Control group:} A control condition in which no intervention-induced changes are expected \textit{a priori}, apart from the intrinsic fluctuations of network activity. These fluctuations are used to characterize the background variability, or noise, against which intervention-induced changes are estimated.

\item \textbf{Reference data:} A dataset or experimental condition providing a reference against which the magnitude of an observed effect can be interpreted. For example, in a dose--response experiment, measurements obtained at different drug concentrations can serve as reference conditions for evaluating the relative magnitude of the response.
\end{enumerate}

Ideally, these three elements should be provided by independent datasets or experimental conditions. However, depending on the experimental design, some of them may be derived from the same dataset or shared across multiple conditions. The framework is therefore designed to accommodate different experimental configurations while maintaining a consistent basis for quantifying and interpreting changes in network activity. See Table \ref{tab:experimental_paradigms} for details regarding datasets roles in the experiments of this work.

\subsubsection{Data Structuring}

Given a multichannel recording of duration $T$, and following the spike detection procedure described above, a spike train $S_{i,n}$ is obtained, where $i=1,2,\ldots,N_{ch}$ denotes the channel index, with $N_{ch}$ representing the total number of recorded channels, and $n$ denotes the index of the detected spike. The recording is divided into $N$ consecutive temporal windows, with the $j$-th window having duration $T_j$.

When an intervention is applied at time $t=t_B$, the first temporal window is defined as the baseline window, with duration $T_B$. The remaining recording is divided into $N$ post-intervention windows, with durations $T_j$, $j=1,2,\ldots,N$. The duration and boundaries of these windows may be selected according to the experimental design, provided that each window contains a sufficient number of spikes to ensure reliable statistical estimation. These post-intervention windows are subsequently used to characterize the temporal evolution of the network response following the intervention. See Table \ref{tab:experimental_paradigms} for data structures involved in the experiments of this work.

\begin{table}[htb!]
    \centering
    \caption{Summary of the experimental procedures, temporal data structure, and analysis strategy for the three experimental paradigms.}
    \label{tab:experimental_paradigms}
    \small
    \renewcommand{\arraystretch}{1.5}
    \setlength{\tabcolsep}{8pt}

        \begin{tabularx}{\textwidth}{
            >{\raggedright\arraybackslash}p{0.16\textwidth}
            @{\hspace{12pt}}
            >{\raggedright\arraybackslash}p{0.23\textwidth}
            @{\hspace{12pt}}
            >{\raggedright\arraybackslash}p{0.27\textwidth}
            @{\hspace{12pt}}
            >{\raggedright\arraybackslash}X
        }
        \toprule
        \textbf{Experiment} &
        \textbf{Experimental groups / conditions} &
        \textbf{Temporal data structure} &
        \textbf{Analysis strategy} \\
        \midrule

        \textbf{Pharmacodynamics}
        \textbf{(4-AP)}
        &
        \begin{enumerate}[leftmargin=*]
            \item 10~$\mu$M 4-AP
            \item 100~$\mu$M 4-AP
            \item H$_2$O (vehicle control)
            \item Medium (control)
        \end{enumerate}
        
        &
        Baseline: 10~min of spontaneous activity. Post-intervention: 30--60~min of recording, divided into 5-min temporal windows.
        &
        \scriptsize The response to 4-AP was quantified relative to baseline activity. Control conditions were used to define the thresholds for significant changes. Responses obtained at 10 and 100~$\mu$M 4-AP were compared to characterize the effect of drug concentration.
        \\

        \textbf{Optogenetic response}
        &
        \begin{enumerate}[leftmargin=*]
            \item ArchT + light stimulation
            \item Light stimulation without ArchT
            \item Spontaneous activity
        \end{enumerate}

        &
        Baseline: 10~min of spontaneous activity.
        \newline Post-intervention: 30--60~min of recording.
        \newline Optogenetic stimulation consisted of repeated 15-s illumination followed by 60-s recovery periods.
        &
        \scriptsize Light-induced suppression of activity in ArchT-expressing cultures was characterized relative to baseline. Cultures without ArchT expression served as a control for light-induced effects. The post-stimulation rebound response was evaluated in relation to the baseline activity.
        \\

        \textbf{Phenotypic response (BIC benchmark)} \cite{crocco2025proper}
        &
        \begin{enumerate}[leftmargin=*]
        \item CYC
        \item 50--50
        \item 80--20
        \item SAG
        \end{enumerate}

        &
        Baseline: 6~min of spontaneous activity. Post-intervention: 6~min of recording, divided into 5-min temporal windows.
        &
        \scriptsize The response to bicuculline was compared across cultures with different inhibitory phenotypes. No vehicle control was available; therefore, spontaneous activity was used as the reference condition for assessing intervention-induced changes.
        \\

        \bottomrule
    \end{tabularx}
\end{table}

\subsubsection{Network Discriminators}

A \textit{discriminator} is defined here as any measurable parameter of network activity derived from the electrophysiological data that can be used to characterize changes in network state following an intervention. Discriminators may capture different aspects of neuronal activity, including its magnitude, temporal structure, or spatial distribution across the recording channels.

The basic discriminator introduced in this work is the mean firing rate (MFR), defined for the $j$-th temporal window as
\begin{equation}
\mathrm{MFR}_j=\frac{\sum_{i,n}S_{i,n}^{(j)}}{N_{ch}T_j},
\end{equation}
where $S_{i,n}^{(j)}$ denotes the $n$-th detected spike on channel $i$ within the $j$-th temporal window, $N_{ch}$ is the number of recorded channels, and $T_j$ is the duration of the corresponding window. The MFR therefore represents the average number of detected spikes per channel per unit time within each temporal window.

We further extend the analysis by introducing additional spiking discriminators based on statistical properties of the \textit{interspike interval} (ISI) probability density function. These discriminators provide complementary information on the temporal organization of network activity and are described in the following subsections.

\subsubsection{Change Metrics}

To quantify changes in a discriminator $\xi$ between two temporal windows, a normalized change metric is required. Several approaches can be used, including the absolute difference $\xi^{(j)}-\xi^{(b)}$, the relative change $(\xi^{(j)}-\xi^{(b)})/\xi^{(b)}$, and ratio-based measures. While these metrics can be useful depending on the application, reference-normalized measures become sensitive to the choice of reference and may become unstable when $\xi^{(b)}$ is small. Here, $j$ denotes an arbitrary temporal window of the recording, while $b$ denotes the baseline window.

Here, we use a symmetric difference-over-sum normalization, also referred to as a normalized contrast, defined as
\begin{equation}
\Delta_{\xi} =
\frac{\xi^{(j)}-\xi^{(b)}}{\xi^{(j)}+\xi^{(b)}}.
\label{eq:contrast}
\end{equation}
This metric is independent of the absolute scale of $\xi$ and treats the two compared values symmetrically. For non-negative discriminators, $\Delta_{\xi}$ ranges from $-1$ to $1$, with $\Delta_{\xi}=0$ indicating no change and the sign indicating the direction of the change. The difference-over-sum formulation has been used in several fields as a normalized measure of relative differences and has also been applied to quantify changes in neuronal response properties \cite{black2017spontaneous}.




\subsection{\textit{Interspike Interval-} Based Analysis}

In this section, we systematically describe the step-by-step implementation of the proposed method for inter-spike interval (ISI) analysis, a fundamental metric for characterizing neuronal firing dynamics in microelectrode array (MEA) recordings.



\subsubsection{Kernel Density Estimation of ISI Distributions}

The ISI is defined as the temporal interval between two consecutive spikes in a spike train. Given a sequence of spikes $S=\{s_1,s_2,\dots,s_N\}$ with corresponding timestamps $\{t_{s_1},t_{s_2},\dots,t_{s_N}\}$, the $n$-th interspike interval is given by:

\begin{equation}
    \mathrm{ISI}_n = t{s_{n+1}} - t_{s_n}
\end{equation}

The analysis is centered on the ISI as a random variable, which has been shown in previous studies to exhibit characteristic probability density distributions \cite{pasquale2010self}. In mature cultured neuronal networks, the ISI distribution frequently exhibits a bimodal structure reflecting two distinct temporal scales: an intra-burst interval, typically ranging from a few milliseconds to a few tens of milliseconds, and an inter-burst interval, typically on the order of hundreds of milliseconds. Such behavior is commonly associated with synchronized network dynamics. Under different experimental conditions, however, alternative distributional profiles may emerge.

To characterize these distributions quantitatively, we estimated the ISI probability density function (PDF) for each recording channel. Since ISI values typically span several orders of magnitude, the data were first transformed using the logarithmic variable $X=\log_{10}(\mathrm{ISI})$. Histograms of $X$ were then constructed for each MEA recording channel using a bin width of $\delta X = 0.1$ \cite{pasquale2010self}. The logarithmic representation improves the interpretability of the distributions and facilitates the approximation of the underlying continuous density from discretized histogram data. The resulting histograms were subsequently used to estimate a continuous probability density function via kernel density estimation (KDE) with an Epanechnikov kernel \cite{epanechnikov1969non}, yielding an empirical approximation of the distribution, denoted by $f_X$.

Consequently, each culture was represented by a set of logarithmic ISI PDFs, $\{f_{X_i}\}_{i=1}^{N_{ch}}$, where $N_{ch}$ denotes the number of active recording channels (electrodes) in the culture. Variations in network dynamics were subsequently assessed through changes in these distributions over time. To this end, the PDFs were estimated within predefined temporal windows $T_j$, each corresponding to a specific experimental phase, including baseline and intervention periods of fixed duration.

\subsubsection{ISI PDF Discriminators}

We derived a set of statistical descriptors from the estimated ISI probability density functions $f_{X_i}^{(j)}$, where $i$ denotes the recording channel index and $j$ the temporal window index.

\begin{itemize}

    \item \textbf{Expected ISI value}:  
    A characteristic timescale of the firing activity was estimated through the expected value (statistical mean) of the logarithmic ISI distribution. Since the PDFs were estimated in the logarithmic domain, the mean was computed as the first moment of $f_{X_i}^{(j)}$:

    \begin{equation}
        \log_{10}\left(\mu_{i,j}\right)
        =
        \int_{-\infty}^{+\infty}
        x f_{X_i}^{(j)}(x)\,dx
    \end{equation}
    
    where $\mu_{i,j}$ denotes the characteristic ISI associated with channel $i$ during temporal window $j$. This parameter provides a global estimate of the dominant temporal scale of neuronal firing activity. In the limiting case of a narrowly distributed or approximately stationary ISI distribution, the average firing rate (AFR) can be approximated as the inverse of the characteristic ISI, i.e.,
\begin{equation}
    \mathrm{AFR} \approx \frac{1}{\mu}
\end{equation}

    \item \textbf{Peak ISI value}:  
    In addition to the mean value, we estimated the mode of the ISI distribution, corresponding to the most probable interspike interval. Since neuronal ISI PDFs commonly exhibit well-defined modal structures, particularly in synchronized bursting networks, the peak location was defined as:

    \begin{equation}
        \log_{10}\left(p_{i,j}\right)
        =
        \underset{x}{\arg\max}
        \left(
        f_{X_i}^{(j)}(x)
        \right)
    \end{equation}

    where $p_{i,j}$ represents the ISI value associated with the maximum of the estimated PDF. Unlike the expected value, which reflects the overall distribution shape, the peak value is primarily sensitive to the dominant firing regime. For approximately symmetric unimodal distributions, one expects $p_{i,j}\approx\mu_{i,j}$, whereas deviations between the two quantities may indicate skewness or multimodality in the firing dynamics.

\end{itemize}





\subsubsection{Kullback–Leibler Divergence}
The Kullback--Leibler divergence, $D_{KL}$, is a statistical metric that quantifies the difference between two probability density functions. Since the mean or the mode of an ISI distribution can remain unchanged while its modality or overall shape changes, $D_{KL}$ provides a robust measure of distributional changes over time \cite{sabottke2013method}. However, although $D_{KL}$ captures the extent of these changes, it does not reveal their direction or physiological interpretation, such as whether they reflect faster or slower neuronal dynamics. It is worth noting that $D_{KL}$ inherently quantifies the difference between two probability distributions and therefore provides an alternative measure of change between datasets. In this sense, $D_{KL}$ can serve as an alternative to $\Delta_{\xi}$ when the discriminator is represented by a probability distribution rather than a scalar quantity.

\subsection{Compositional Analysis of Network Response States}

Here, we extend the analysis by introducing a finer temporal decomposition of each time window $j$ of length $T_j$ to characterize the response of the culture and resolve the network state transitions induced by the intervention. Each temporal window is further divided into $K$ smaller segments of length $\tau$ ($T_j=K\cdot\tau$), each containing a sufficient number of spikes to estimate the logarithmic ISI probability density function $f_X$. In principle, each temporal window $j$ represents a characteristic response time scale, e.g., minutes for drug-induced responses, whereas the segments within each window provide individual samples of the network response over shorter time intervals. For each segment, the response of each electrode is characterized using an arbitrary discriminator $\xi$ and quantified relative to its baseline value through the contrast index $\Delta_\xi^{(i,j,k)}$ (Eq.~\ref{eq:contrast}), where $i$ denotes the electrode, $j$ the temporal window, and $k$ the segment. The electrode-level responses are then classified according to the magnitude and direction of the change, corresponding to increased, decreased, or unchanged activity. Each temporal window $j$ is subsequently characterized by the network state resulting from the composition of these electrode-level responses across its constituent segments.

\begin{figure}[]
    \centering
    \includegraphics[width=\linewidth]{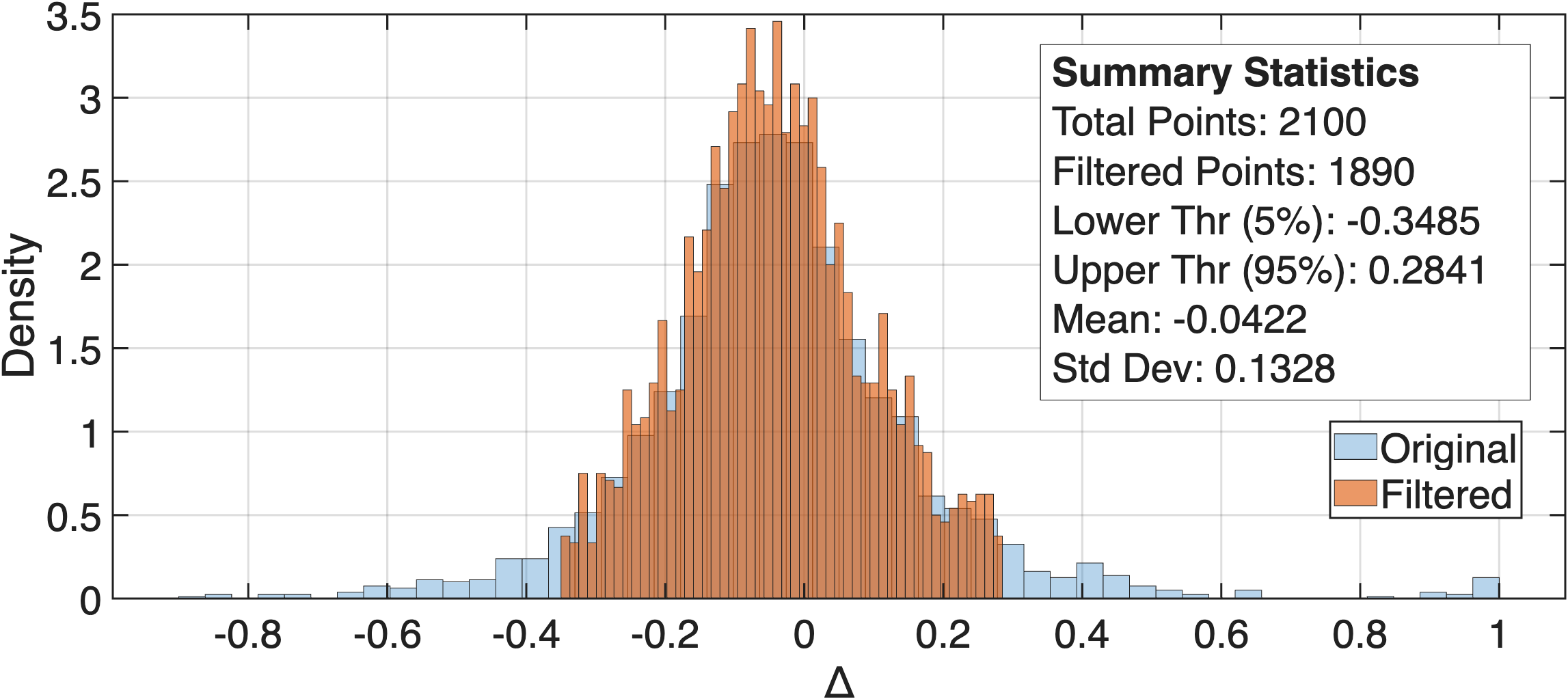}
    \caption{\textbf{Determination of the significance threshold for changes in a given discriminator from vehicle-control recordings.} Empirical distribution of the contrast index ($\Delta$) derived from vehicle-only experiments. The dashed lines indicate the 5th and 95th percentiles of the empirical control distribution, used as the lower and upper response thresholds, respectively. In this example the analysis is shown for mean ISI, $\mu_{\mathrm{ISI}}$.}
    \label{fig:water_threshold}
\end{figure}

To define the levels of electrode response and distinguish intervention-specific effects from spontaneous network variability, a controlled statistical framework was implemented. For each intervention, a corresponding vehicle control condition was defined, such as the water group when the compound was dissolved in water. Control cultures were exposed exclusively to the vehicle, and the resulting $\Delta_\xi$ values were used to characterize the range of natural network variability. These values were pooled to construct an empirical distribution representing intrinsic network fluctuations under control conditions. Depending on the experimental design and the intervention, the control distribution and corresponding thresholds can be determined over the entire recording duration $T$ or separately for each temporal window $T_j$. The latter approach may be appropriate when the vehicle condition itself produces time-dependent changes, for example, due to fatigue during prolonged recordings. The resulting distribution serves as a data-driven null model of natural network variability. Thresholds were defined from the 95\% and 5\% confidence limits of the control distribution~\cite{herrera1958precision}, yielding upper and lower bounds, $\Delta_{th\uparrow}$ and $\Delta_{th\downarrow}$, respectively (Fig. \ref{fig:water_threshold}). These thresholds were subsequently applied to the intervention groups to identify electrode-level responses that deviated from the range of variability observed under control conditions.

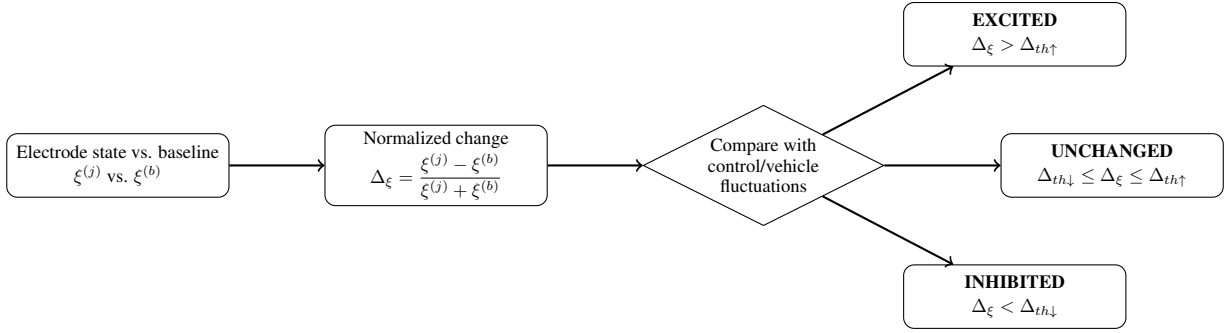
\begin{figure}[]
\centering
\begin{tikzpicture}[scale=0.7, transform shape,
    node distance=1.2cm and 1.8cm,
    box/.style={
        draw,
        rounded corners,
        align=center,
        minimum width=4.2cm,
        minimum height=1.2cm
    },
    decision/.style={
        draw,
        diamond,
        align=center,
        aspect=2,
        inner sep=2pt
    },
    arrow/.style={->, thick}
]

\node[box] (state) {
    Electrode state vs. baseline\\[2pt]
    $\xi^{(j)}$ vs. $\xi^{(b)}$
};

\node[box, right=of state] (contrast) {
    Normalized change\\[2pt]
    $\displaystyle
    \Delta_\xi =
    \frac{\xi^{(j)}-\xi^{(b)}}
         {\xi^{(j)}+\xi^{(b)}}$
};

\node[decision, right=of contrast] (control) {
    Compare with\\
    control/vehicle\\
    fluctuations
};

\draw[arrow] (state) -- (contrast);
\draw[arrow] (contrast) -- (control);

\node[box, above right=1.3cm and 1.5cm of control] (excited) {
    \textbf{EXCITED}\\[2pt]
    $\Delta_\xi > \Delta_{th\uparrow}$
};

\node[box, right=2.2cm of control] (unchanged) {
    \textbf{UNCHANGED}\\[2pt]
    $\Delta_{th\downarrow}
    \leq \Delta_\xi
    \leq \Delta_{th\uparrow}$
};

\node[box, below right=1.3cm and 1.5cm of control] (inhibited) {
    \textbf{INHIBITED}\\[2pt]
    $\Delta_\xi < \Delta_{th\downarrow}$
};

\draw[arrow] (control) -- (excited);
\draw[arrow] (control) -- (unchanged);
\draw[arrow] (control) -- (inhibited);

\end{tikzpicture}
\caption{
Classification of electrode-wise responses based on the normalized change
$\Delta_\xi$ relative to the natural fluctuations observed in the
control or vehicle condition.
}
\label{fig:response_classification}
\end{figure}

Following the intervention, the response of each neuronal culture was characterized using the discriminator $\xi$ over $N$ consecutive temporal windows, each of duration $T_j$. Each window $j$ was further divided into segments $k$ of duration $\tau$. For each electrode $i$ and segment $k$, the response was classified as excited ($E$), inhibited ($I$), or unchanged ($U$) according to the corresponding value of the discriminator $\Delta_\xi^{(i,j,k)}$ and the predefined upper and lower thresholds, $\Delta_{th\uparrow}$ and $\Delta_{th\downarrow}$ (see also Fig. \ref{fig:response_classification}):

\begin{equation}
    \gamma \left(\Delta_\xi^{(i,j,k)} \right)\in
    \left\{
    \begin{array}{lll}
        E & \text{if } \Delta_\xi^{(i,j,k)}>\Delta_{th\uparrow}:
          & \textbf{Excited activity state}\\
        I & \text{if } \Delta_\xi^{(i,j,k)}<\Delta_{th\downarrow}:
          & \textbf{Inhibited activity state}\\
        U & \text{if } \Delta_{th\downarrow}\leq \Delta_\xi^{(i,j,k)}\leq\Delta_{th\uparrow}:
          & \textbf{Unchanged activity state}
    \end{array}
    \right.
\end{equation}

Each temporal window $j$ was subsequently represented, for each electrode $i$, by a three-component state vector
$\mathbf{s}^{(i,j)}=(s_E^{(i,j)},s_I^{(i,j)},s_U^{(i,j)})$, where each component quantifies the fraction of temporal segments associated with the corresponding response state $E,I,U$. Specifically, for $\gamma\in{E,I,U}$, we define

\begin{equation}
    s_\gamma^{(i,j)}=\frac{\sum_{k=1}^K \mathbb{I}_\gamma \left ( \Delta_\xi^{(i,j,k)} \right) \cdot \left | \Delta_\xi^{(i,j,k)} \right |}
    {\sum_{k=1}^K \left | \Delta_\xi^{(i,j,k)} \right |}
\end{equation}

where $\mathbb{I}_\gamma(\cdot)$ is the indicator function given by
\begin{equation}    
\mathbb{I}_\gamma(x) = 
\begin{cases}
1 & \text{if} \quad x\in \gamma \\
0 & \text{otherwise}
\end{cases}
\end{equation}

Thus,

\begin{equation}
    \sum_{\gamma\in{E,I,U}} s^{(i,j)}_\gamma = 1.
\end{equation}

\begin{figure}
    \centering
    \includegraphics[width=0.75\linewidth]{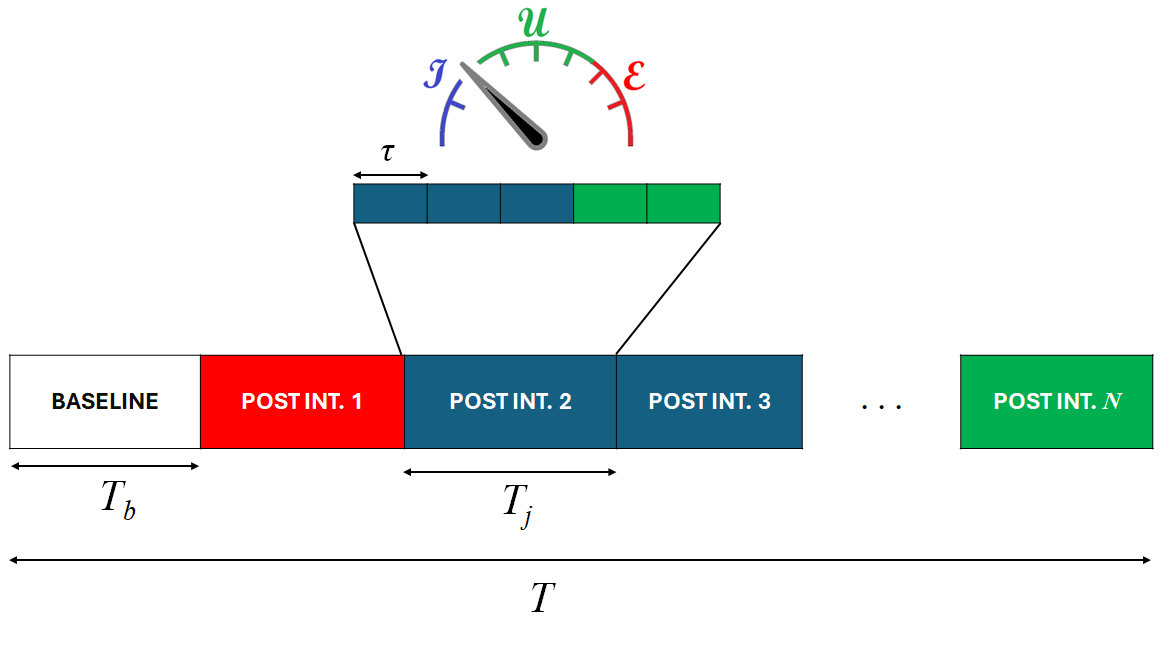}
    \caption{\textbf{Description of the compositional analysis:} Each post-intervention window $j$ is further segmented into shorter temporal segments. In each segment, all active electrodes are classified into one of three response states, as described above and in Figure \ref{fig:response_classification}: excited (E), inhibited (I), or unchanged (U). The resulting distribution of electrode states defines the compositional state vector, illustrated in the figure as a gauge representing its relative composition.}

    \label{fig:Compositional_description}
\end{figure}

Together, these three components provide a compact representation of the electrode-specific response composition within each temporal window, capturing the relative persistence of the excited, inhibited, and unchanged states following the intervention (Fig. \ref{fig:Compositional_description}).

\begin{figure}
    \centering
    \includegraphics[width=0.95\linewidth]{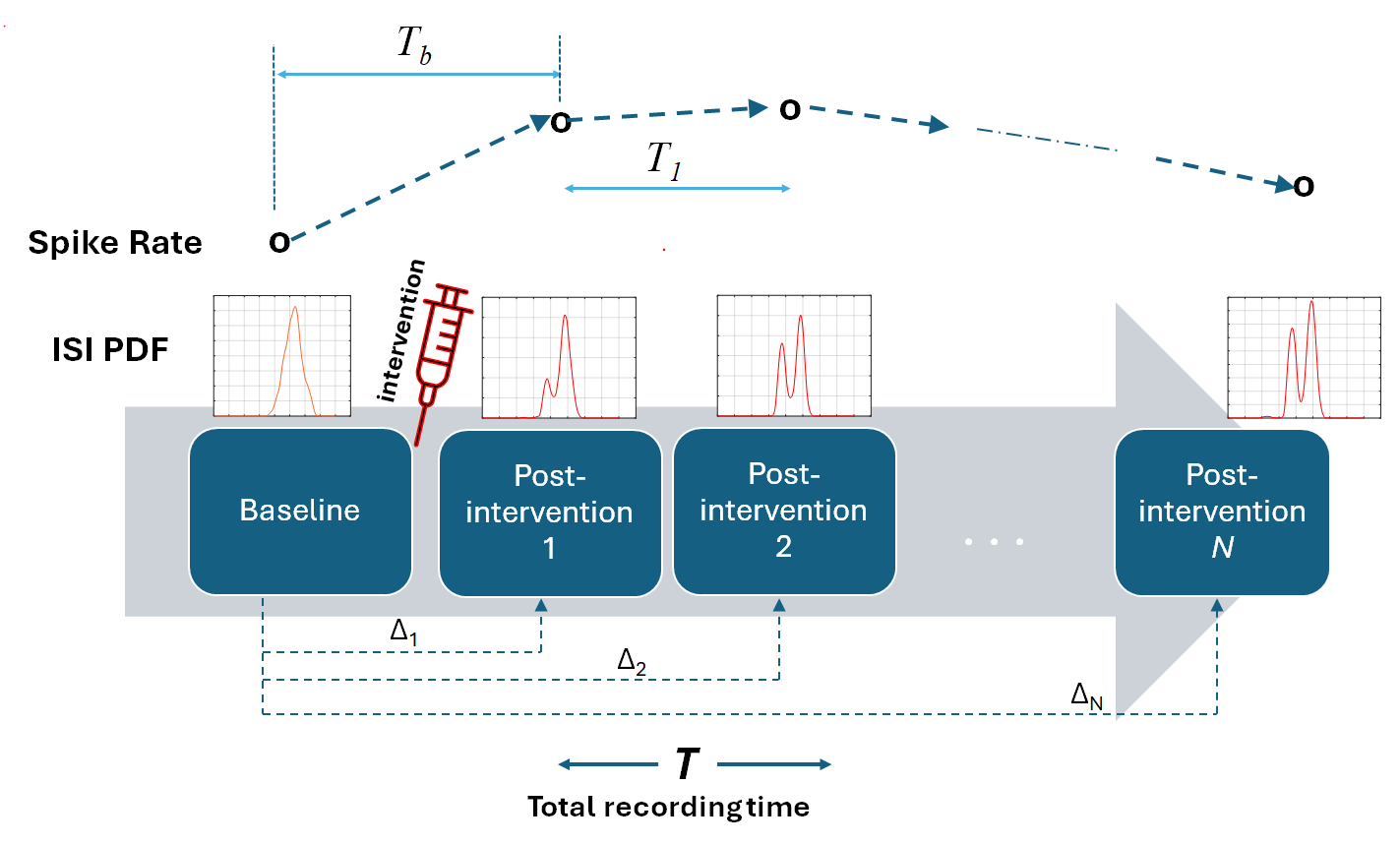}
    \caption{\textbf{Conceptual schematic of the experimental and data-analysis pipeline:} An electrophysiological recording of total duration $T$ is divided into a pre-intervention baseline period of duration $T_b$ and a post-intervention period. The latter is further divided into $N$ temporal windows of duration $T_j$, each of which is compared with the baseline window using a measurable discriminator of network activity. Changes in the discriminator between the baseline and post-intervention windows are quantified using the metric $\Delta$ (Eq. \eqref{eq:contrast} in this work). The discriminator may be based on the mean firing rate (MFR) or on statistics derived from the interspike interval (ISI).}

    \label{fig:exp_General}
\end{figure}

\begin{figure}[htbp]
    \centering
    \includegraphics[width=0.85\textwidth]{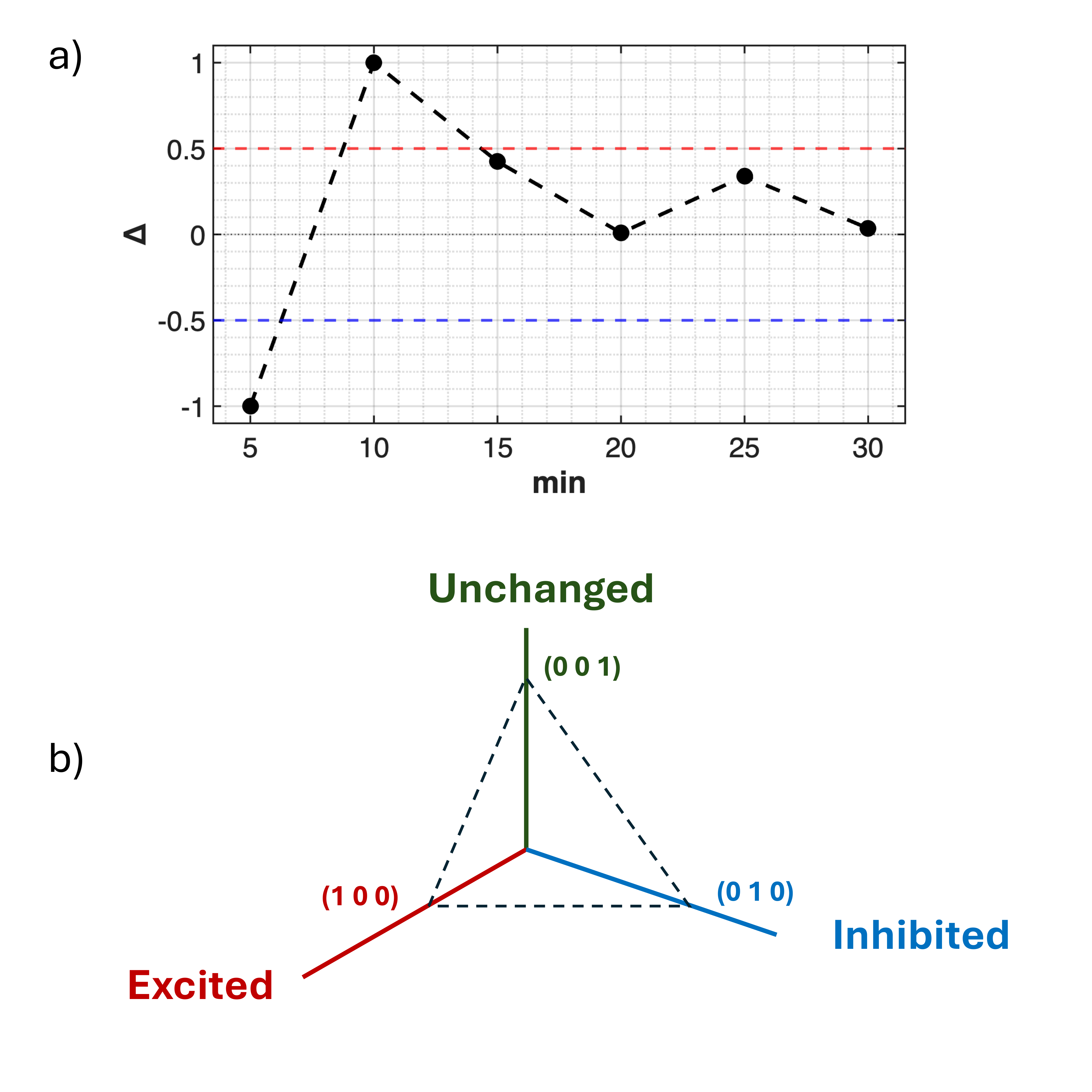}
    \caption{\textbf{Temporal activity classification and representation within the excitation--inhibition--unchanged (EIU) space}. 
    (a) Example of experimental data from a single MEA electrode, showing the normalized mean inter-spike interval (ISI) over time. The upper and lower thresholds, derived from control recordings, define the boundaries used to classify temporal changes in electrode activity. 
    (b) Three-dimensional EIU space representing the relative contributions of excited, inhibited, and unchanged activity. Each electrode is mapped within this space according to its computed excitation, inhibition, and unchanged state $(s_E,s_I,s_U)$, such that its position reflects the relative contribution of each activity state over time.}
    \label{fig:EIU_space}
\end{figure}

The response of an entire neuronal culture was summarized by averaging the state vectors across all recording electrodes:

\begin{equation}
\bar{\mathbf{s}}_j=\frac{1}{N_{ch}}\sum_{i=1}^{N_{ch}}\mathbf{s}^{(i,j)} \stackrel{\mathrm{def}}= \{\mathcal{E}_j,\mathcal{I}_j,\mathcal{U}_j\},
\end{equation}

where $N_{ch}$ is the number of recording electrodes and $\mathbf{s}^{(i)}$ denotes the state vector of electrode $i$. Since $\mathcal{E}+\mathcal{I}+\mathcal{U}=1$, $\mathbf{s}$ is a compositional vector that lie on a two-dimensional simplex. Each vertex of the simplex corresponds to a pure dynamical state (100\% excited, depressed, or unchanged), while interior points represent mixtures of these states~\cite{aitchison1982statistical}.

For visualization, the simplex was projected onto an equilateral triangle using the barycentric transformation

\begin{align}
X &= \mathcal{E} + \frac{1}{2}\mathcal{U},\\
Y &= \frac{\sqrt{3}}{2}\mathcal{U}.
\end{align}

This transformation preserves the proportional relationships while enabling clear 2D visualization \cite{aitchison1982statistical}.

From the transformation into the excitation–inhibition-unchanged (EIU) state space, two quantities with potential biological interpretation can be extracted:

\begin{equation}
    \mathcal{R} = \mathcal{E} + \mathcal{I},
\end{equation}

where $\mathcal{R}$ represents the overall level of network activation, and

\begin{equation}
    \eta = \log \left( \frac{\mathcal{E}}{\mathcal{I}} \right),
\end{equation}

where $\eta$ is the E-I balance, which characterizes the relative dominance between excitatory and inhibitory activity.

These measures may provide a quantitative framework for characterizing pharmacological responses across different classes of drugs and may further assist in identifying optimal dosage ranges that ensure both efficacy and safety in drug development.

\subsection{ Immunohistochemistry assessments}
Fixed neurons were permeabilized with 0.1\% Triton X-100 in phosphate-buffered saline (PBS) for 20 minutes at room temperature. To reduce nonspecific antibody binding, cells were subsequently blocked with 3\% bovine serum albumin (BSA) in PBS for 1 hour. Neurons were then incubated overnight at 4°C with primary antibodies diluted in 3\% BSA: mouse anti-NeuN (1:500, Millipore) and rabbit anti-Parvalbumin (1:1000, Swant).

Following primary antibody incubation, samples were washed several times with PBS containing 0.1\% BSA and incubated for 2 hours at room temperature with fluorophore-conjugated secondary antibodies. Alexa Fluor 568–conjugated anti-mouse IgG was used to visualize NeuN (red), and Alexa Fluor 488–conjugated anti-rabbit IgG was used to detect Parvalbumin (green) \cite{zaccaria2026investigation}.

After secondary incubation, cells were washed thoroughly with PBS and counterstained with DAPI (Invitrogen) to label nuclei. Samples were then mounted using Aqua Polymount (Polysciences Inc.) and stored at 4°C until imaging. Fluorescence images were acquired with appropriate filter sets to detect DAPI, Alexa Fluor 488, and Alexa Fluor 568 signals.

\subsection{Statistical analysis}

Statistical analyses were performed using MATLAB R2022b. Data distributions were assessed for normality using the Shapiro–Wilk test \cite{ghasemi2012normality}. For normally distributed data, two-group comparisons were conducted using independent-samples \textit{t}-tests, and comparisons across three or more groups were performed using one-way analysis of variance (ANOVA), followed by Tukey’s honestly significant difference (HSD) post hoc test when appropriate \cite{agbangba2024use}. 

For non-normally distributed data, two-group comparisons were performed using the Mann–Whitney \textit{U} test, paired comparisons using the Wilcoxon signed-rank test, and multi-group comparisons using the Kruskal–Wallis test, followed by Dunn’s post hoc correction where applicable \cite{chicco2025simple}. 

All statistical tests were two-sided. Statistical significance was defined as $p < 0.05$, unless otherwise specified.

\section{Results}

\subsection{Aggregate Analysis of Network Activity}

\begin{figure*}[h]
    \centering
    \includegraphics[width=\textwidth]{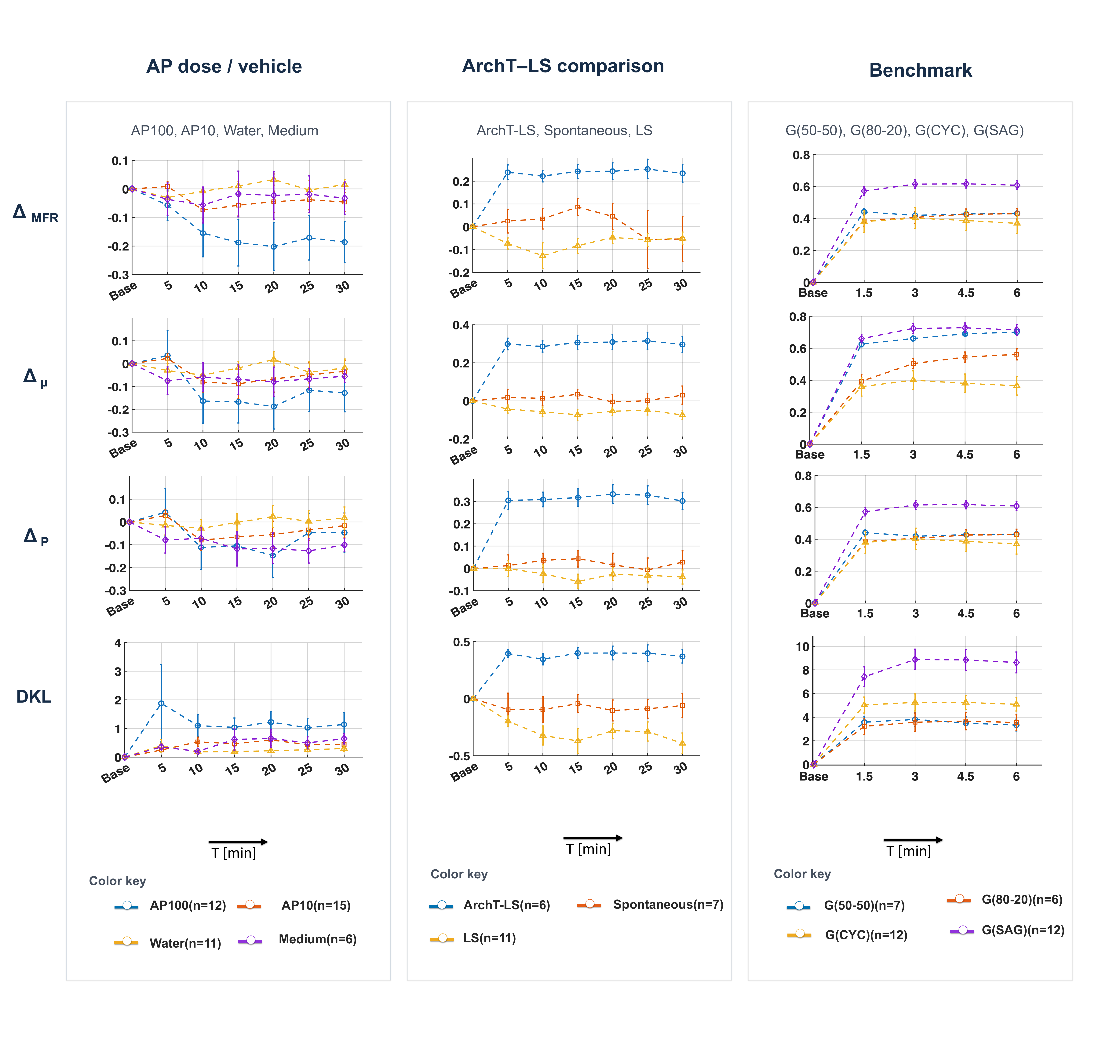}
    \caption{\textbf{Time-dependent response profiles across experimental paradigms.}
    Comparative overview of the temporal response patterns observed across the different experimental paradigms. The reorganized representation highlights the evolution of the measured responses over time and facilitates direct comparison among experimental conditions, emphasizing both shared temporal trends and paradigm-specific differences.Data are presented as group means ± SEM, where n represents the number of independent MEA cultures analyzed per experimental group.}
    \label{fig:time_dependent_response_profiles}
\end{figure*}

In the aggregate analysis, changes in a given discriminator $\xi$ were tracked over time by calculating the contrast $\Delta_\xi$ (Eq.~\eqref{eq:contrast}) for each electrode and temporal window $j$. The resulting electrode-level values were then averaged across electrodes to obtain a single measure of network change for each culture. Culture-level responses were subsequently aggregated within each experimental group, and their statistical significance was assessed by comparison with the corresponding control group. Table~\ref{tab:discriminators_explanation} summarizes the discriminators considered in the analysis. Figure~\ref{fig:time_dependent_response_profiles} presents the corresponding time-dependent response profiles for each experimental group and discriminator.

\begin{table}[h]
    \centering
    \caption{Network discriminators considered in the analysis and their interpretation in terms of neuronal network activity.}
    \label{tab:discriminators_explanation}
    \small
    \renewcommand{\arraystretch}{1.25}
    \setlength{\tabcolsep}{5pt}
    
    \begin{tabularx}{\columnwidth}{
        >{\raggedright\arraybackslash}p{0.30\columnwidth}
        >{\raggedright\arraybackslash}X
    }
        \toprule
        \textbf{Discriminator} ($\xi$) &
        \textbf{Interpretation} \\
        \midrule

        Mean firing rate (MFR)
        &
        Average number of detected spikes per unit time and per active recording channel within the considered temporal window. Changes in MFR indicate changes in the overall level of network spiking activity and may be associated with increased or decreased network excitability.
        \\[5pt]

        Mean ISI ($\mu$)
        &
        Mean value of the interspike-interval distribution. Changes in $\mu$ reflect changes in the characteristic temporal separation between consecutive spikes and, consequently, in the overall temporal organization of network activity.
        \\[5pt]

        Peak of ISI PDF ($p$)
        &
        Location of the dominant mode of the interspike-interval distribution. Changes in $p$ indicate a shift in the characteristic spiking interval and may reflect alterations in the temporal coordination or dynamical regime of the network.
        \\[5pt]

        Kullback--Leibler divergence ($D_{KL}$)
        &
        Quantifies the difference between two ISI probability distributions. Larger values indicate greater divergence in the temporal distribution of interspike intervals and, consequently, a stronger change in spiking dynamics between the compared conditions.
        \\

        \bottomrule
    \end{tabularx}
\end{table}



\subsection{Static EIU compositional analysis}

To characterize the overall response of neuronal cultures following intervention, we first applied the EIU framework as a static compositional representation. For each culture, activity states classified over the complete 30-min recording period were summarized by their relative contributions to the excited (E), inhibited (I), and unchanged (U) states. The resulting EIU composition therefore represents the overall direction and prevalence of activity changes across the observation period, rather than their temporal evolution.

\begin{figure*}[htbp]
    \centering
    \includegraphics[width=\textwidth]{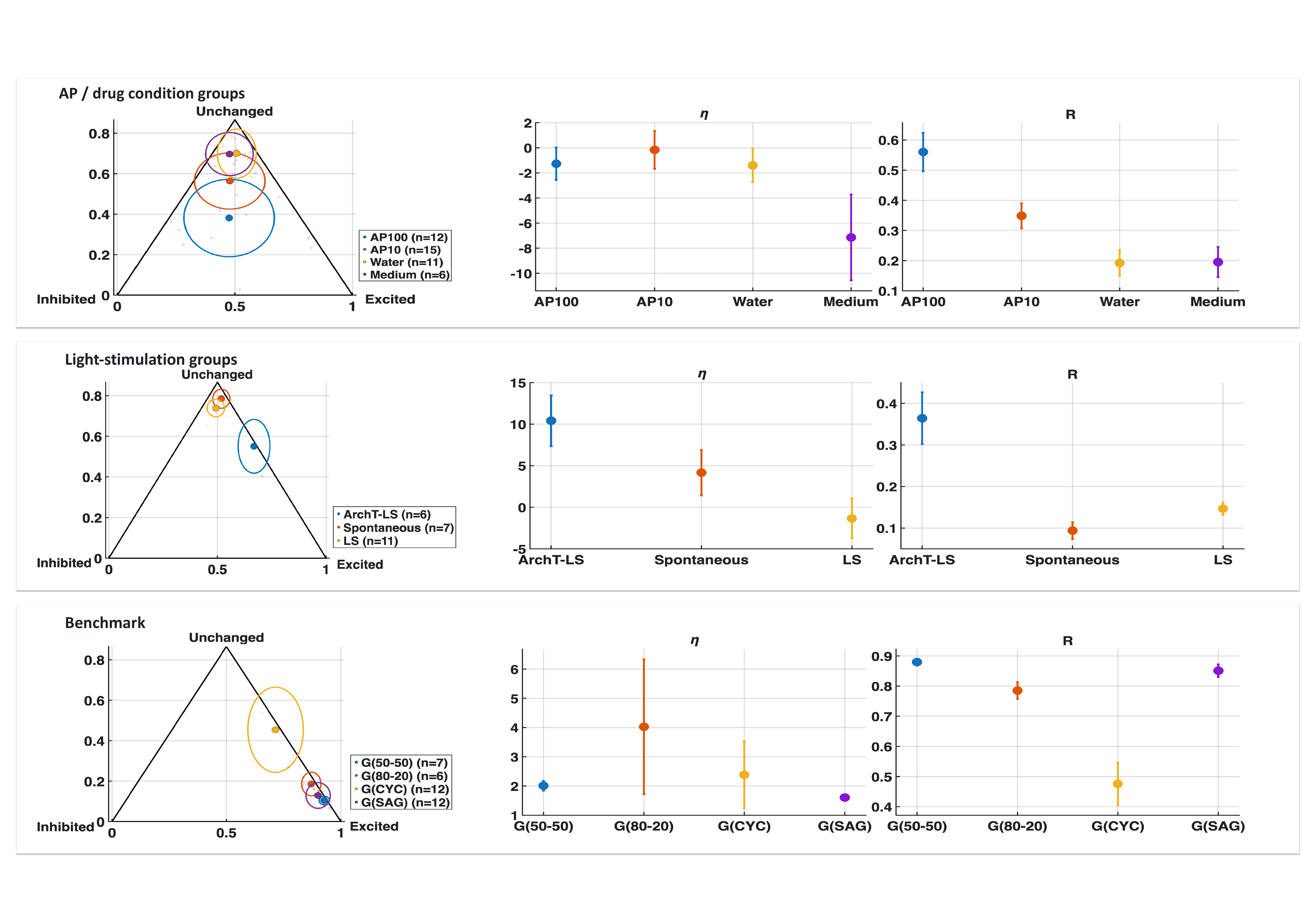}
    \caption{\textbf{Static EIU composition and group-level metrics across experimental conditions.}
    The left column shows the distribution of each experimental group within the excited--inhibited--unchanged (EIU) activity-state simplex, where position reflects the relative contributions of faster-spiking (E), slower-spiking (I), and unchanged (U) activity states across the complete observation period.
    The middle and right columns show the corresponding response-direction balance ($\eta$) and responsiveness ($R$), respectively.
    Points represent group means and error bars indicate the standard error of the mean (SEM). The sample size (n) indicated for each condition represents the number of independent MEA cultures included in the analysis.
    Rows correspond to the 4-AP intervention, optogenetic stimulation, and external benchmark datasets. Statistical significance was defined as $p<0.05$.}
    \label{fig:static_EIU_composition}
\end{figure*}

The responsiveness metric ($R$) ,  Figure~\ref{fig:static_EIU_composition} AP/ drug condition, provided and important information by quantifying the overall prevalence of activity changes irrespective of their direction. AP100 exhibited the greatest overall responsiveness among the chemical-intervention conditions, whereas AP10 showed a more moderate response and the control conditions remained comparatively less responsive. Thus, the static EIU representation distinguished both the magnitude of the network response, through $R$, and its predominant direction, through $\eta$.

A distinct response profile was observed in the optogenetic paradigm (Figure~\ref{fig:static_EIU_composition}, second row). The ArchT-LS group exhibited the greatest responsiveness among the optogenetic conditions, indicating a larger overall deviation from baseline activity compared with the spontaneous and light-control groups. In addition, the displacement of the ArchT-LS group toward the excited activity-state direction, together with its corresponding $\eta$, indicated a predominance of faster-spiking activity relative to baseline. In contrast, the spontaneous and light-control conditions showed substantially smaller overall responses.

Finally, we evaluated the static EIU representation using the external benchmark dataset, in which four culture compositions were exposed to the same bicuculline intervention (Figure~\ref{fig:static_EIU_composition}, third row). The different culture compositions occupied distinct positions within the EIU simplex and showed corresponding differences in $R$ and $\eta$. These results indicate that, under the same pharmacological perturbation, the EIU representation can distinguish response profiles associated with different culture compositions.

Taken together, the static analysis shows that $R$ and $\eta$ provide complementary descriptions of the network response. Whereas $R$ quantifies the overall prevalence of intervention-associated activity changes, $\eta$ describes the predominant direction of those changes toward faster- or slower-spiking activity states. However, because the static representation summarizes activity across the complete observation period, it does not reveal when these activity-state transitions occur. We therefore next applied the EIU framework independently at successive time points to characterize the temporal trajectories underlying the static response profiles.

\subsection{Time-resolved EIU compositional analysis}

To investigate how the EIU response evolved over time, the compositional transformation was calculated independently at successive time points throughout the post-intervention recording period. In contrast to the static representation, which summarizes the complete observation period as a single composition, the time-resolved analysis tracks the trajectory of each experimental group through the EIU activity-state space.

\begin{figure*}[htbp]
    \centering
    \includegraphics[width=\textwidth]{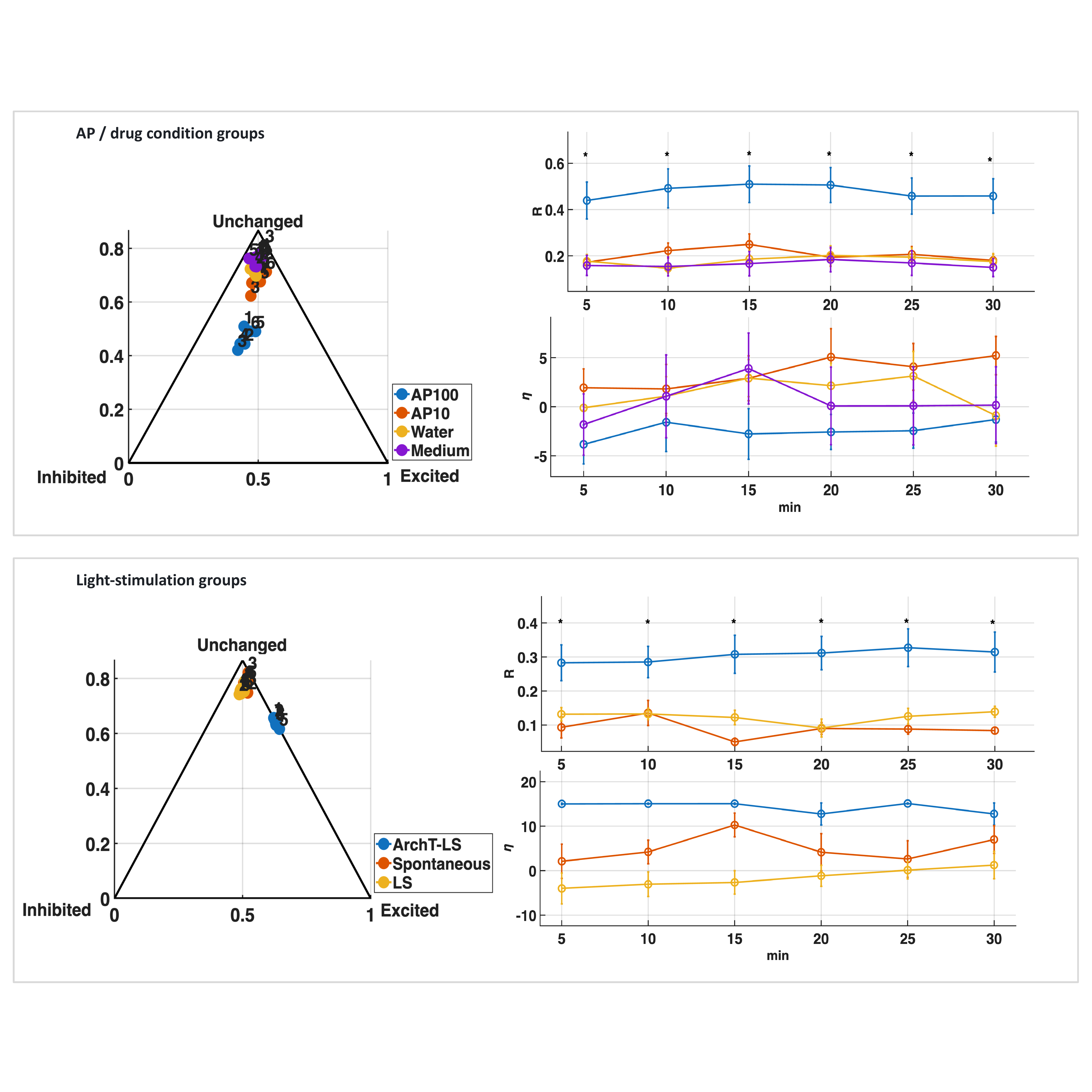}
    \caption{\textbf{Time-resolved EIU composition and group-level metrics.}
    Temporal evolution of network activity within the excited--inhibited--unchanged (EIU) activity-state framework across experimental conditions.
    The EIU simplex plots show the trajectories of each group over successive time points, illustrating changes in the relative contributions of faster-spiking (E), slower-spiking (I), and unchanged (U) activity states.
    The corresponding time-resolved metrics, $\eta$ and $R$, quantify response-direction balance and overall network responsiveness, respectively.
    The upper panel presents the 4-AP/drug-condition groups (AP100, AP10, Water, and Medium), whereas the lower panel presents the light-stimulation groups (ArchT-LS, Spontaneous, and LS).
    Together, these representations illustrate both the direction of activity-state transitions within the EIU space and the temporal evolution of the associated network-level metrics from 5 to 30~min. Points represent group means and error bars indicate the standard error of the mean (SEM), with n representing the number of independent MEA cultures analyzed per experimental group. Statistical significance was defined as $p<0.05$.}
    \label{fig:timepassing_EIU_composition}
\end{figure*}

For the chemical-intervention conditions, the time-resolved representation revealed how the response profiles observed in the static analysis developed throughout the 30-min recording period (Figure~\ref{fig:timepassing_EIU_composition}). Rather than remaining at a fixed position within the EIU simplex, the treatment groups followed distinct trajectories relative to their baseline states. In particular, the AP100 condition progressively shifted toward the inhibited activity-state direction, consistent with the predominance of slower-spiking activity observed in the static representation. The corresponding temporal profiles of $R$ and $\eta$ further describe the development of the response in terms of its magnitude and direction, respectively.

The optogenetic conditions exhibited a different temporal response pattern. As shown in Figure~\ref{fig:timepassing_EIU_composition}, the ArchT-LS group followed a pronounced trajectory toward the excited activity-state direction, accompanied by an increase in responsiveness. In comparison, the spontaneous and light-control groups showed smaller changes within the EIU space. The time-resolved representation therefore reveals not only the overall response detected by the static analysis but also how the direction and magnitude of that response evolved throughout the observation period.

\section{Discussion}

In this study, we introduce a quantitative analytical framework for characterizing intervention-induced changes in neuronal network activity. Rather than reducing neuronal responses to changes in aggregate activity alone, the proposed EIU framework quantifies alterations in the underlying organization of network activity. We evaluated its performance across neuronal cultures subjected to distinct perturbation paradigms, including acute pharmacological stimulation with 4-aminopyridine (4-AP), prolonged optogenetic inhibition, and an independent inter-laboratory dataset used as a benchmark \cite{crocco2025proper}. Together, these experimental conditions provided a range of response magnitudes and temporal profiles against which the sensitivity and interpretability of the framework could be assessed.

A central limitation of conventional electrophysiological metrics, particularly MFR, is that they compress complex temporal responses into a single measure of activity magnitude \cite{tomar2021variability,wen2021capturing}. Although such metrics remain highly informative for identifying pronounced changes in network excitability, averaging can obscure biologically meaningful differences in the temporal structure of the response \cite{tomar2021variability,wen2021capturing}. For example, a strong but transient initial response may be substantially attenuated when firing activity is averaged over a longer observation window, potentially obscuring short-lived changes in neuronal activity \cite{la2021mean}. In contrast, the EIU framework retains information regarding both the temporal organization and persistence of intervention-induced changes, thereby distinguishing response profiles that may produce similar aggregate activity levels (Figure~\ref{fig:static_EIU_composition}). This distinction is particularly relevant when interventions induce dynamic or nonstationary responses rather than sustained monotonic changes in activity (Figure~\ref{fig:timepassing_EIU_composition}).

\begin{figure}[htbp]
    \centering
    \includegraphics[width=\linewidth]{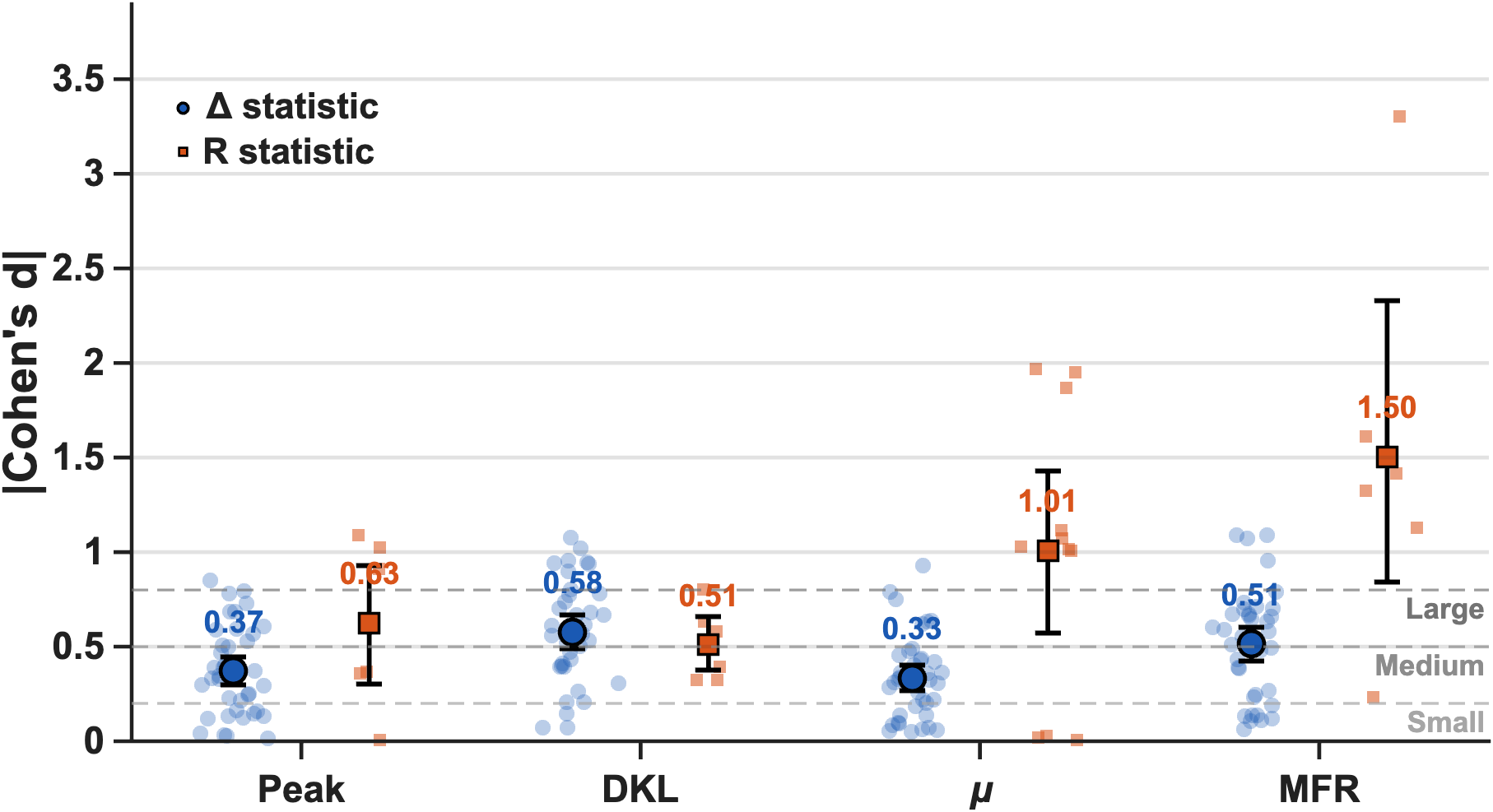}
    \caption{Comparison of effect-size magnitude between $\Delta$ and $R$ statistical approaches across Peak, DKL, $\mu$, and MFR metrics. Individual points represent absolute Cohen's $d$ values, while larger markers indicate the mean and error bars show 95\% bootstrap confidence intervals. Dashed reference lines denote small ($d=0.2$), medium ($d=0.5$), and large ($d=0.8$) effect-size thresholds.}
    \label{fig:cohensD}
\end{figure}

A second limitation of conventional summary metrics becomes apparent when intervention effects are small or moderate. Subtle alterations in spike timing or temporal organization may occur in the absence of substantial changes in average firing rate, making such effects difficult to resolve using rate-based measures alone \cite{vieira2024temporal}. Under such conditions, averaging can reduce the separation between experimental groups, limiting the ability of conventional statistical tests to detect intervention-induced effects \cite{kass2005statistical,tomar2019methods}. This behavior was particularly evident in the 4-AP intervention groups, where relatively modest changes were not consistently resolved using conventional metrics Figure~\ref{fig:time_dependent_response_profiles} . We therefore investigated whether distribution-derived descriptors could increase sensitivity while retaining a direct relationship to established electrophysiological quantities. The shift in the peak of the ISI distribution, for example, provided a more sensitive representation of intervention-induced changes than conventional averaging and produced a modest improvement in effect size, as quantified by Cohen's (d) (Figure~\ref{fig:cohensD}). These findings indicate that incorporating information from the underlying activity distribution can improve the detectability of subtle network responses.

The strongest discrimination, however, was obtained using ($D_{KL})$, which captures changes in the overall shape and modality of the ISI distribution rather than changes in a single summary statistic Figure~\ref{fig:cohensD}. Consequently, ($D_{KL})$ was particularly sensitive to small-to-moderate perturbations that were difficult to distinguish using conventional metrics Figure~\ref{fig:cohensD}. This enhanced sensitivity comes at the cost of directional interpretability: A divergence measure quantifies the magnitude of the difference between two activity distributions but does not, by itself, indicate the physiological direction of the underlying change \cite{lundstrom2006decoding}. Thus, a large divergence establishes that the network state has changed substantially, but does not directly reveal whether neuronal activity has shifted toward faster or slower firing dynamics.

To overcome this limitation, we developed \textbf{EIU compositional analysis}, which integrates the sensitivity of distribution-based divergence with information about the direction and structure of the underlying electrophysiological change. This representation enabled not only discrimination between experimental groups exposed to different intervention strengths or doses, but also characterization of \textbf{how} network activity changed (Figure~\ref{fig:time_dependent_response_profiles}). Importantly, this improvement was supported quantitatively by effect-size analysis. Cohen's (d) increased substantially for several metrics when the EIU framework was applied, with the largest improvements observed for mean ISI and MFR, whereas ($D_{KL})$, which already exhibited relatively strong discrimination, showed no further increase in effect size (Figure~\ref{fig:cohensD}). The framework therefore does not simply generate an alternative visualization of neuronal activity; rather, it increases the statistical separability of network states, particularly when the underlying intervention produces subtle or moderate effects.

Notably, the largest gain in effect size was observed when the EIU framework was applied to MFR (Figure~\ref{fig:cohensD}). This result is important because it indicates that the improvement does not depend exclusively on the choice of a particular electrophysiological feature, such as ISI. Instead, the framework can enhance the information extracted even from a conventional metric such as MFR by changing how intervention-induced deviations are represented and compared. The relative performance of individual input metrics may therefore depend on the biological question and perturbation under investigation, whereas the principal advantage of the EIU framework lies in its ability to amplify otherwise subtle differences between network states.

An important consideration is that the gain in statistical separability provided by the EIU framework was not uniform across all input metrics or effect magnitudes. In particular, when the underlying electrophysiological descriptor already exhibited relatively strong discrimination between experimental conditions, as observed for ($D_{KL})$, transformation into the EIU representation did not further increase the effect size and could produce a comparable or slightly reduced Cohen's (d) (Figure~\ref{fig:cohensD}). One possible explanation is the bounded nature of the compositional representation. Because (E+I+U=1) and the responsiveness metric is defined as (R=E+I), (R) is constrained between 0 and 1. As the proportion of time points classified as responsive approaches unity, increasingly large changes in the underlying electrophysiological signal cannot produce proportional increases in (R). This potential saturation should therefore be considered when interpreting strongly separated conditions. The EIU framework should consequently be viewed primarily as a complementary approach for enhancing the detection and characterization of subtle or moderate, temporally persistent responses, rather than as a transformation expected to increase effect size under all experimental conditions.

Collectively, these findings suggest that the EIU framework occupies a complementary position between conventional summary statistics and purely distribution-based divergence measures. Conventional metrics such as MFR provide intuitive directional information but can lose temporal and distributional structure through averaging, whereas ($D_{KL})$ offers greater sensitivity to changes in network-state distributions but lacks an intrinsic indication of physiological direction \cite{tomar2021variability,wen2021capturing,lundstrom2006decoding}. EIU compositional analysis combines these complementary properties, providing both enhanced sensitivity and interpretable characterization of intervention-induced changes. Across the perturbation paradigms examined here, this advantage was most evident for small-to-moderate effects, for which conventional approaches showed limited statistical separation. The framework may therefore provide a broadly applicable quantitative strategy for detecting and characterizing subtle changes in neuronal network dynamics that would otherwise remain concealed within conventional aggregate measures.

Our analytical approach is capable of enhancing statistical separability, particularly for subtle-to-moderate responses, while simultaneously tracking the temporal trajectory of culture responses following stimulation. By transforming ISI distributions into an EIU compositional simplex representation, this method provides detailed insight into the responsiveness and dynamic balance of neuronal cultures under different stimulation paradigms. Our findings demonstrate that this framework is not only effective for tracing drug-induced or optogenetically evoked responses, but is also sensitive to intrinsic culture-specific properties, thereby enabling a more comprehensive characterization of how neuronal networks adapt to diverse stimuli.

Importantly,the framework may provide a quantitative basis for future studies examining dose-dependent network responses and could potentially complement preclinical efficacy or safety assessments. In addition, the framework can be used to optimize stimulation parameters in optogenetic experiments. For example, in the present study we employed prolonged light-on windows to evaluate ArchT-mediated responses in neuronal cultures. Interestingly, the faster-spiking state observed in the ArchT-LS condition may be compatible with post-illumination rebound firing previously reported in ArchT-expressing neurons; however, identifying the underlying mechanism was beyond the scope of the present computational study\cite{chuong2014noninvasive,higgs2021periodic}.
Depending on the experimental objective, stimulation duration and frequency can be systematically adjusted while continuously tracking culture trajectories, thereby enabling optimization of the desired functional outcome.

 The benchmark dataset provided an important complementary validation of the proposed framework. In contrast to the more subtle responses observed in the other experimental paradigms, bicuculline induced pronounced changes across all culture compositions \cite{crocco2025proper}. Under these conditions, where the experimental groups were already well separated by conventional statistical measures, a substantial further increase in effect size following transformation was not necessarily expected. Nevertheless, the benchmark dataset revealed an important feature of the network response that extended beyond statistical separability. Across $\mu$, $D_{KL}$, and the other evaluated metrics, the responses increased rapidly following bicuculline exposure and subsequently approached relatively stable plateaus Figure~\ref{fig:time_dependent_response_profiles}. Importantly, although all culture groups responded strongly to the same pharmacological perturbation, they converged toward different response plateaus, with the magnitude of these responses varying according to culture composition Figure~\ref{fig:time_dependent_response_profiles}. This observation indicates that the effect of a common perturbation cannot be characterized solely by whether a response occurs or by its statistical significance; the resulting response state also reflects differences among the underlying neuronal populations \cite{crocco2025proper}. Thus, the benchmark analysis demonstrates that the value of the proposed framework is not restricted to enhancing the detection of subtle effects, but also extends to characterizing distinct network response states when intervention-induced changes are already pronounced.

In this study, 4-AP was expected to increase neuronal activity; however, the observed response was characterized by a reduction in activity, contrary to the anticipated excitatory effect \cite{dietrich2021neuroprotective,jackson2007state}. Only a modest initial change in activity was detected, followed by a reduction in MFR during continued 4-AP exposure see Figure~\ref{fig:time_dependent_response_profiles}. One possible interpretation is that sustained exposure to 4-AP induced a progressive reduction in network responsiveness, producing a desensitization-like pattern of activity rather than persistent excitation \cite{veres2024cellular,rienecker2020merits}. In this scenario, prolonged 4-AP-induced hyperexcitability may recruit homeostatic mechanisms that progressively suppress or stabilize neuronal firing, thereby contributing to the observed reduction in MFR \cite{michetti2024homeostatic}. However, because the underlying cellular mechanisms were not directly investigated, our data cannot distinguish pharmacological desensitization from other activity-dependent or homeostatic adaptations.

\begin{figure}[htbp]
    \centering
    \includegraphics[width=\linewidth]{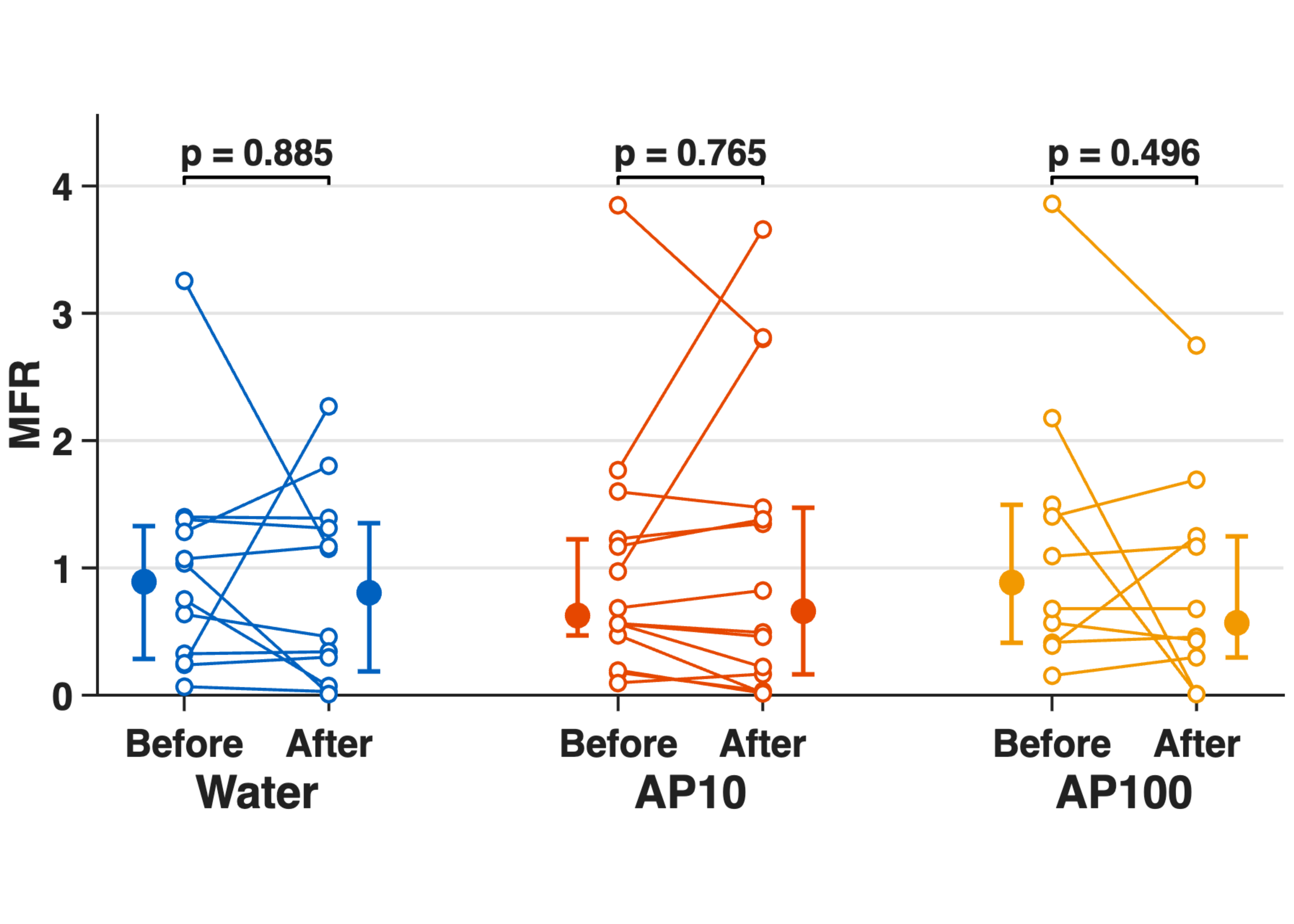}
    \caption{\textbf{Neuronal activity 24 h after 4-AP removal and characterization of the neuronal culture.}
\textbf{} Cultures were exposed to 4-AP for 30 min, after which the drug was removed and the cultures were washed. Mean firing rate (MFR) was reassessed 24 h later to evaluate whether neuronal activity persisted following 4-AP removal. Water-treated cultures served as controls. Individual paired measurements are connected by lines, showing the change in MFR between the initial (Before) and 24-h follow-up (After) recordings. Each connected pair represents one independent MEA culture measured before and 24 h after treatment. Statistical comparisons were performed using the Wilcoxon signed-rank test; exact $p$-values are indicated above each group. The persistence of neuronal activity 24 h after 4-AP removal is consistent with the reduced activity observed following 4-AP exposure reflecting a prolonged adaptive or desensitizing response rather than a complete loss of neuronal activity.
}
    \label{fig:DAYAFTER}
\end{figure}

To determine whether this unexpected reduction reflected a persistent alteration in network function, cultures were reassessed 24 h after 4-AP removal (Fig.~\ref{fig:DAYAFTER}). MFR was no longer significantly different from pre-exposure levels, indicating recovery of network activity and suggesting that the effect was reversible. This recovery is consistent with a transient adaptive or desensitization-like response to 4-AP rather than persistent impairment of neuronal function \cite{michetti2024homeostatic}.

\section*{Conclusion}

In conclusion, we developed a time-resolved EIU compositional framework for characterizing intervention-induced changes in neuronal network activity beyond conventional aggregate electrophysiological measures. By representing activity as excited, inhibited, and unchanged states relative to baseline, the framework provides complementary information on the magnitude, direction, and temporal evolution of network responses. Across pharmacological, optogenetic, and independent benchmark datasets, the greatest advantage of the EIU framework was observed for small-to-moderate responses, where conventional measures showed limited separation between experimental conditions and the compositional transformation enhanced statistical discriminability. In contrast, when the underlying response was already strongly distinguishishable, the framework provided additional value by characterizing response direction, temporal evolution, and differences in the resulting network states rather than necessarily further increasing effect size. Together, these findings position EIU analysis as a complementary approach for detecting subtle neuronal network responses while providing an interpretable representation of how network states evolve following perturbation.
\newpage
\section*{Acknowledgments}
This work was financed by the European Union - NextGenerationEU and National Recovery and Resilience Plan (NRRP) - Mission 4 Component 2 Investment 1.2 – “Funding projects presented by young researchers” MSCA PNRR Young Researchers, “CIRCUS project” - MSCA20240000106 - CUP E63C25000820007.

We would like to thank Federico Cremisi from bio@SNS lab and collaborators from Scuola Normale Superiore in Pisa, Italy for sharing with us the datasets of bicuculline response used to benchmark the method presented in this paper. We would like also to thank Martina Grandi for helping us in developing the spike detection code segment.

\clearpage
\bibliographystyle{ieeetr}  
\bibliography{references}

\end{document}